\documentclass[sigconf]{acmart}
\usepackage{CJKutf8}
\usepackage{amsthm}
\newtheorem{defn}{Definition}

\usepackage{makecell}
\usepackage{threeparttable}
\usepackage{tabularx}
\usepackage{subcaption}
\usepackage{graphicx}

\AtBeginDocument{%
  }

\setcopyright{acmlicensed}
\copyrightyear{2018}
\acmYear{2018}
\acmDOI{XXXXXXX.XXXXXXX}
\acmConference[Conference acronym 'XX]{Make sure to enter the correct
  conference title from your rights confirmation email}{June 03--05,
  2018}{Woodstock, NY}
\acmISBN{978-1-4503-XXXX-X/2018/06}

\begin{document}
\begin{CJK*}{UTF8}{bsmi}

\title{CSRM-LLM: Embracing Multilingual LLMs for Cold-Start Relevance Matching in Emerging E-commerce Markets}

\settopmatter{authorsperrow=5}
\author{Yujing Wang}
\author{Yiren Chen}
\authornote{Corresponding author. Email: yichen24@coupang.com}
\author{Huoran Li}
\affiliation{%
  \institution{Coupang, Inc.}
  \state{Beijing}
  \country{China}
}

\author{Chunxu Xu}
\author{Yuchong Luo}
\author{Xianghui Mao}
\affiliation{%
  \institution{Coupang, Inc.}
  \state{Beijing}
  \country{China}
}

\author{Cong Li}
\author{Lun Du}
\author{Chunyang Ma}
\affiliation{%
  \institution{Coupang, Inc.}
  \state{Beijing}
  \country{China}
}

\author{Qiqi Jiang}
\author{Yin Wang}
\author{Fan Gao}
\affiliation{%
  \institution{Coupang, Inc.}
  \state{Beijing}
  \country{China}
}

\author{Wenting Mo}
\author{Pei Wen}
\affiliation{%
  \institution{Coupang, Inc.}
  \state{Beijing}
  \country{China}
}

\author{Shantanu Kumar}
\affiliation{%
  \institution{Coupang, Inc.}
  \state{Seoul}
  \country{Republic of Korea}
}

\author{Taejin Park}
\author{Yiwei Song}
\affiliation{%
  \institution{Coupang, Inc.}
  \state{Mountain View}
  \country{United States}
}

\author{Vijay Rajaram}
\author{Tao Cheng}
\affiliation{%
  \institution{Coupang, Inc.}
  \state{Mountain View}
  \country{United States}
}

\author{Sonu Durgia}
\author{Pranam Kolari}
\affiliation{%
  \institution{Coupang, Inc.}
  \state{Mountain View}
  \country{United States}
}
\renewcommand{\shortauthors}{Yujing Wang, Yiren Chen, Huoran Li, et al.}

\begin{abstract}
As global e-commerce platforms continue to expand, companies are entering new markets where they encounter cold-start challenges due to limited human labels and user behaviors. In this paper, we share our experiences in Coupang to provide a competitive cold-start performance of relevance matching for emerging e-commerce markets. Specifically, we present a \textbf{C}old-\textbf{S}tart \textbf{R}elevance \textbf{M}atching (CSRM) framework, utilizing a multilingual Large Language Model (LLM) to address three challenges: (1) activating cross-lingual transfer learning abilities of LLMs through machine translation tasks; (2) enhancing query understanding and incorporating e-commerce knowledge by retrieval-based query augmentation; (3) mitigating the impact of training label errors through a multi-round self-distillation training strategy. Our experiments demonstrate the effectiveness of CSRM-LLM and the proposed techniques, resulting in successful real-world deployment and significant online gains, with a 45.8\% reduction in defect ratio and a 0.866\% uplift in session purchase rate.
\end{abstract}

\maketitle

\section{Introduction}
Large cross-border e-commerce platforms are experiencing rapid growth, offering consumers convenient access to products from a wide range of regions. As companies expand into new markets, they encounter cold-start challenges stemming from limited human-labeled data and insufficient user behavior signals, which hinder the optimization of relevance models for the target language. Providing accurate relevance matching performance on day one is crucial for maintaining user engagement and satisfaction. Traditionally, developing customized relevance models for each emerging market was a resource-intensive process, requiring extensive label collection and training the model from scratch. Moreover, ensuring consistency among human annotators by adhering to detailed relevance labeling guidelines is challenging, resulting in noisy training labels. 

With the rapid advancement of large language models (LLMs), many language-dependent machine learning tasks can now be addressed with minimal, or even no, task-specific training data. For the relevance matching task, Microsoft found that their LLM-based labeling method outperformed human raters~\cite{thomas2024large}, while Google focused on enhancing zero-shot LLM rankers by scoring fine-grained relevance labels~\cite{zhuang2024beyond}. 
Although large language models have been extensively explored for relevance prediction, the specific challenges associated with query-product relevance in the e-commerce domain remain underexplored. Relevance modeling in e-commerce requires domain-specific labeling criteria, which limits the effectiveness of off-the-shelf large language models in zero-shot settings. For example, when users search for ``red shirts", they typically expect apparel for humans rather than clothing designed for pets. 
Similarly, when users look for ``Nike sneakers", products from comparable brands like Adidas should be considered acceptable substitutes rather than irrelevant products.
The inherent complexity of the e-commerce relevance assessment makes supervised fine-tuning (SFT) an essential step to fulfill customized experiences.

In this paper, we delve into the challenges of e-commerce relevance matching for cross-language cold-starts within a supervised fine-tuning framework. Specifically, we divide the problem into two stages. The first stage is to initiate a relevance LLM without any labels or user behaviors in the target market. This setting reflects the day-one performance we can provide when first entering a new region. The second stage involves localizing the LLM using sparse data. As relevant labels and user behavior data gradually accumulate in the new market, the objective is to further enhance relevance matching performance by incorporating localized knowledge.

In addressing the above issue, we are confronted with three major challenges:
\begin{enumerate}
    \item \textbf{Cross-Lingual Generalization Gap:} Despite the robust zero-shot capabilities of large language models, there remains a significant semantic disparity between the training samples in the source language and the evaluating samples in the target language. A key issue lies in how to activate the cross-lingual generalization potential of LLMs for the cold-start relevance matching problem.
    \item \textbf{Domain-Specific Query Ambiguity:} User queries tend to be short and may carry unique meanings in the local e-commerce context. For example, the query term ``神仙水" (``miracle water") commonly refers to a specific SK-II skincare product rather than a literal beverage. 
    Incorporating specific e-commerce knowledge is essential for enhancing query understanding and relevance matching.
    \item \textbf{Label Noise from Human Annotation:} When collecting local relevance labels for the new market, human annotators often fail to get fine-grained alignment with the labeling guidelines. 
    Therefore, devising a strategy to minimize the adverse effects of noisy training data becomes imperative.
\end{enumerate}


    

To tackle these challenges, we propose a \textbf{C}old-\textbf{S}tart \textbf{R}elevance \textbf{M}atching (CSRM) framework that leverages multilingual LLMs for relevance prediction in emerging e-commerce markets. For the first challenge, we begin by collecting human-translated queries, titles, brand names and categories from the source to the target language. Then, we jointly optimize relevance matching and machine translation within a unified LLM. By learning from both tasks, the LLM is better equipped to perform relevance matching in the target language. 
For the second problem, we enrich the e-commerce knowledge by retrieving semantically similar product titles. While Retrieval-Augmented Generation (RAG)~\cite{lewis2020retrieval} has shown effectiveness across tasks, cold-start scenarios lack behavioral data for training retrieval models. We design a novel Retrieval-based Query Augmentation (RQA) approach that leverages LLMs to generate pseudo behavioral data for distilling an embedding-based retrieval model. 
To mitigate the impact of training label errors (\textit{i.e.}, the third challenge), we introduce a multi-round self-distillation training strategy. The model is gradually optimized by several rounds of training, while the model from the previous round serves as the teacher to generate soft labels for the subsequent round. Equipped with the self-distillation approach, the noisy training data can be rectified by the soft labels, consequently improving the LLM performance while preventing severe over-fitting. 

Extensive experiments and analyses demonstrated the superiority of CSRM framework and the effectiveness of proposed techniques. To serve an effective relevance model online, we leverage the CSRM-LLM as the teacher model to distill an efficient online model based on a twin-towered architecture~\cite{liu2021que2search}. The model shows superior performance in an online A/B test and has been launched to production, achieving 45.8\% reduction in defect ratio and 0.866\% enhancement in session purchase rate. 

The contributions of this paper are summarized as follows:
\begin{itemize}
    \item We introduce the methodology of a cross-border e-commerce company to roll out relevance models for emerging markets. We show that a multi-lingual LLM is particularly beneficial in the cold-start scenario, while adding auxiliary machine translation tasks in the multi-task fine-tuning stage further activates the cross-lingual generalization ability of LLMs.
    \item We propose a novel method for Retrieval-based Query Augmentation (RQA) when the behavioral training data in the new market is unavailable. The retrieved titles serve as query augmentation for the LLM input, enabling explicit integration of the e-commerce knowledge.
    \item We leverage the multi-round self-distillation training framework to improve LLM generalization. With the self-distillation technique, the LLMs benefit from domain-oriented training data while minimizing the impact of labeling errors.
\end{itemize}

\section{Problem and Feature Definition}

In this section, we introduce the problem definition of relevance matching and the selected features. 

\begin{defn}[\textbf{\emph{Definition 1: Relevance Matching}}]
\label{def:relevance_prediction}
	Relevance matching is a task to learn a conditional probability $\mathbf{\mathcal{P}}(\mathbf{y}|\mathbf{i}, \mathbf{q}; \Theta)$ to distinguish the relevance level of each query $q$ and item $i$ pair, according to a labeled dataset $\mathbf{D}$, where $\Theta$ is the model parameters, $\mathbf{i}$ is the $i$-th item with item side features, $\mathbf{q}$ is the $q$-th query with query side features, and $\mathbf{y} \in \left \{1, 2, \cdots k\right\}$ is the label of the relevance level.
    Here $k$ represents the granularity of level division.
\end{defn}

The features used for relevance modeling are defined in Table \ref{table:feature}. In our scenario, the relevance labels are divided into three levels:  \textbf{(i) Exact match}, where the product exactly matches the search intent (e.g., query is ``sneakers", product is a Nike sneaker); \textbf{(ii) Substitute}, where the product can serve as a substitute (e.g., query is ``Nike shoes", product is an Adidas sneaker); and \textbf{(iii) Irrelevant}, where the product cannot serve as an alternative due to different functionality or non-negotiable attributes (e.g., query is ``woman shoes", product is a woman T-shirt).


Note that we have a detailed labeling guideline for the human raters to reference, which is omitted here for business reasons. However, it is difficult for raters to align with all detailed guidelines, leading to inevitable labeling errors in the training data.


Furthermore, we divide the problem into two stages according to the availability of relevance labels in the target market:
\begin{itemize}
    \item \textbf{Stage-1: Cold-start with zero data}. Given only the relevance labels from established markets, our goal is to develop a cold-start relevance LLM that maximizes the performance for a new market using a different language. 
    \item \textbf{Stage-2: Localization by sparse data}: Once a certain volume of relevance labels and user behavior data have been collected in the new market, the objective of this scenario is to further optimize the relevance performance by leveraging the region-specific knowledge.
\end{itemize}

\begin{table}[t]
\small
\renewcommand\arraystretch{0.95}
\caption{Feature Definitions}
\label{table:feature}
    \begin{threeparttable}
        \begin{tabularx}{0.48\textwidth}{lX} \hline
        \textbf{Feature name} & \textbf{Definition} \\ \hline
        Query & User input text in the search box. \\ \hline
        Query brand & The brand name appearing in the user query, matched by a pre-defined dictionary. \\ \hline
        Product title & Title in target language of the new market \\ \hline
        Product description & The text description of each product \\ \hline
        Product brand & The product brand name given by merchants \\ \hline
        Product category & The product category in the catalog taxonomy given by merchants \\ \hline
        RQA & The retrieval-based query augmentation  generated by an EBR algorithm. This feature will be explained in detail in the methodology section. \\ \hline
        \end{tabularx}
    \end{threeparttable}
\end{table}

\section{Methodology}

\subsection{Overview}

\begin{figure}[h]
  \centering
  \includegraphics[width=1.0\linewidth]{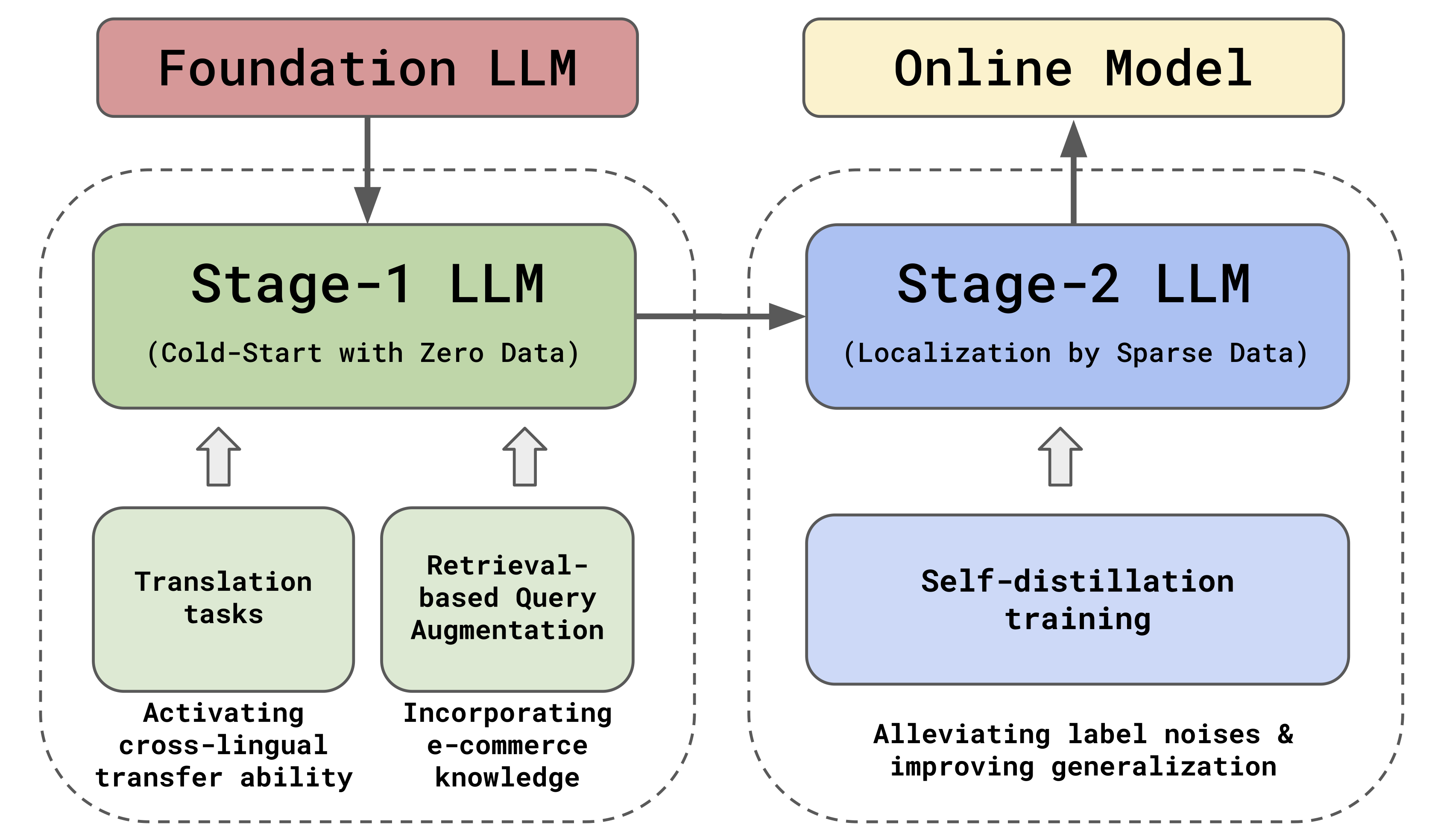}
  \caption{The Overall Framework of CSRM}
  \label{fig:llm_arch}
\end{figure}

Upon receiving a prompt input consisting of a query and product pair, the CSRM-LLM generates one of the three relevance labels: exact match, substitute, or irrelevant.
The overall framework of CSRM are illustrated in Figure \ref{fig:llm_arch}. Starting from an open-sourced foundation model, we perform Low-Rank Adaptation (LoRA) fine-tuning using the relevance labels in the source market, resulting in the Stage-1 LLM. In this stage, we do not utilize the localized labels and user behaviors in the target market, so it reflects the day-one performance we can provide when first entering a new market. To activate the cross-lingual generalization ability and incorporate e-commerce knowledge, we add machine translation as an auxiliary task and propose retrieval-based query augmentation, which effectively improve the performance of the Stage-1 model. Following that, we leverage manual relevance labels in the target market to fine-tune a Stage-2 LLM. To mitigate human labeling errors, we adopt a multi-round self-distillation training strategy to get the final model, CSRM-LLM. The model effectively incorporates local knowledge and reaches a higher relevance score. Finally, we transfer its superior offline performance to online setting through knowledge distillation.  

\subsection{Auxiliary Translation Tasks}
\label{sec:translation}
To leverage the cross-lingual generalization capabilities of large language models (LLMs) for e-commerce relevance, a common approach is to translate training data from the source to the target language using standard machine translation tools. However, our preliminary analysis shows that generic translators (e.g., Google Translate, GPT-4) introduce domain-specific errors in over 30\% of query-product pairs, degrading model performance. To address this, we propose an alternative method that jointly trains LLMs on machine translation and relevance matching tasks. This multi-task setup enables the model to acquire both linguistic and domain-specific knowledge, enhancing relevance prediction in the target language.
In our experiments, Korean is used as the source language and traditional Chinese as the target. To ensure translation quality and domain relevance, training data for the machine translation task is manually curated by professional annotators with expertise in both e-commerce and the respective languages. To jointly train the relevance and auxiliary translation tasks, data samples are interleaved according to predefined ratios and randomly drawn during training to form mini-batches.

\subsection{Retrieval-based Query Augmentation}
\label{sec:rqa}
Retrieval-Augmented Generation (RAG)~\cite{lewis2020retrieval} is a popular method for integrating domain knowledge into Large Language Models (LLMs). For e-commerce relevance matching, we focus on retrieval-based augmentation on the query side, as the user queries are often short and have large semantic gaps with the product-side features.

While a well-trained Embedding-Based Retrieval (EBR) model can effectively extract domain-specific knowledge from external sources, our cold-start scenario lacks sufficient user behavior data to support such training. To address this issue, we employ the term-based BM25 algorithm~\cite{robertson2009probabilistic} to generate pseudo query-product pairs. However, BM25 retrieval introduces considerable noise, degrading training quality and model performance. To mitigate this, we introduce an auxiliary task in CSRM-LLM as the LLM ranker that estimates the likelihood of a product being selected for conversion under a given query. These probabilities serve as soft labels, enabling knowledge distillation into a high-performing EBR model for Retrieval-based Query Augmentation (RQA). The distillation loss combines contrastive loss and mean squared error:

\begin{equation}
\begin{aligned}
    Loss_{cl} & =  \sum_{q} \frac{CVR(q, p) exp(<embed(q), embed_(p)>)}{\sum_{p^-} exp(<embed(q), embed(p^-)>)} \\
    Loss_{mse} & = \sum_{q} \sum_{p} ||CVR(q, p), <embed(q), embed(p)>|| \\
\end{aligned}
\end{equation}
where $Loss_{cl}$ and $Loss_{mse}$ denote the contrasting loss and mean-squared error loss, respectively; $CVR(q, p)$ is the predicted conversion rate of product $p$ under query $q$ by the LLM teacher; $p^-$ denotes a randomly sampled negative product; $<.>$ represents for the inner product function. By emphasizing higher-ranking query-product pairs, we reduce the impact of noisy data and enhance the quality of RQA features. Leveraging the cross-lingual capabilities of LLMs, the soft labels remain effective even when the LLM ranker is trained solely on historical conversions from existing markets and applied directly to new ones.

\subsection{Multi-Round Self-Distillation}
\label{sec:sd}

\begin{figure}[t]
  \centering
  \includegraphics[width=1.0\linewidth]{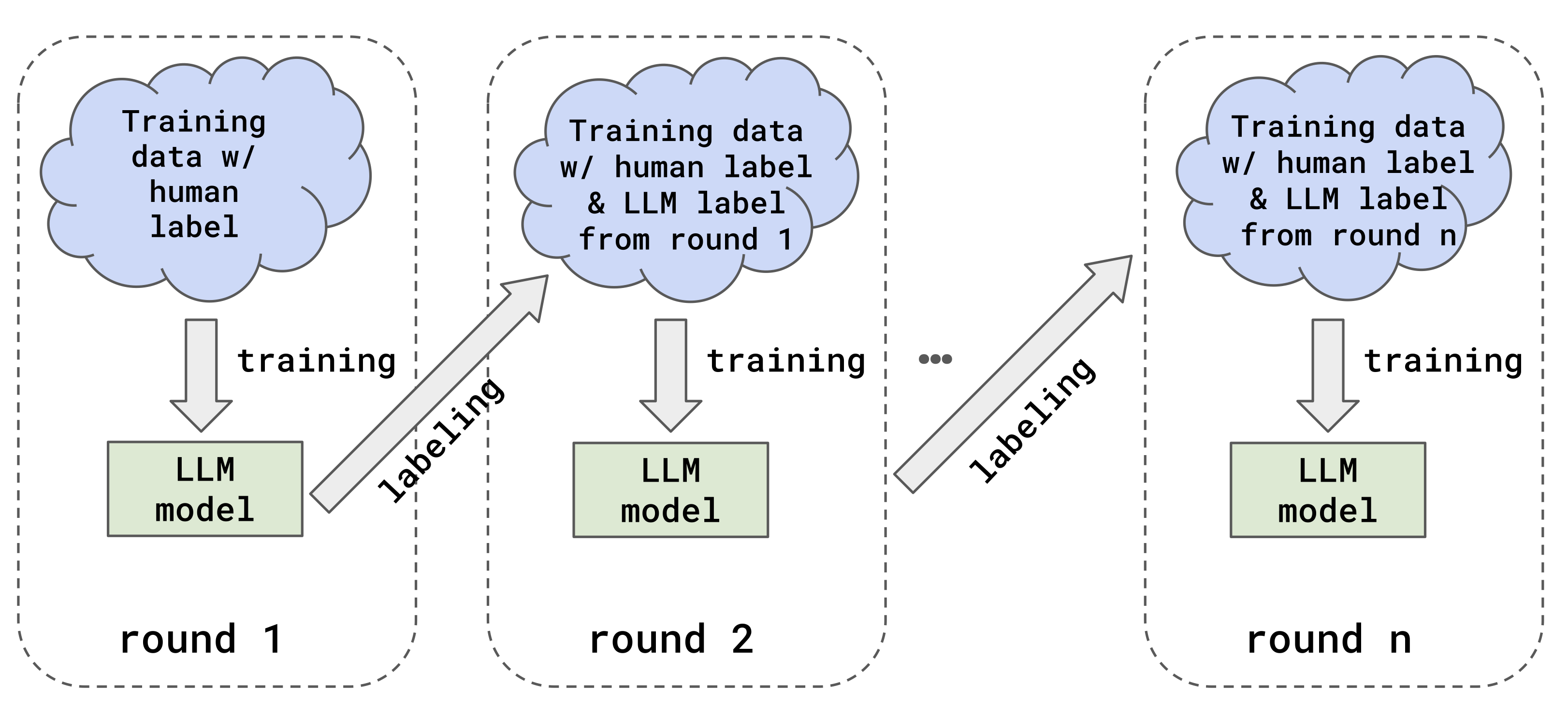}
  \caption{Multi-round Self-distillation Training}
  \label{fig:multi_round_sd}
\end{figure}

As illustrated in Figure \ref{fig:multi_round_sd}, self-distillation proceeds in an iterative manner. In each round, the LLM generates soft labels for the training data used in the subsequent round. The process begins with a model trained solely on human-annotated hard labels and gradually reduces the impact of noisy labels by incorporating soft labels from previous iterations. Each self-distillation cycle consists of two alternating stages, a labeling stage and a training stage. During labeling, the LLM checkpoint from the previous round serves as the teacher model and is used to infer soft labels for all training samples in the next round. The model then infers soft labels for all training samples. Each soft label is represented as a triplet $<p_e, p_s, p_i>$, where $p_e$, $p_s$ and $p_i$ represent the predicted probabilities of the sample to be exact match, substitute and irrelevant, respectively, subjecting to $p_e + p_s + p_i = 1$. In the training process, we adopt the following loss function for LLM optimization:

\begin{equation}
    \sum_{j \in \mathbf{D}} \alpha * CE_{Loss}(hl_j, \hat{p_j}) + (1 - \alpha) * CE_{Loss}(sl_j, \hat{p_j})
\end{equation}
where $hl_j$ and $sl_j$ denote the soft label and human label for the $i$-th sample, respectively. $hl_j = <p_{e,j}, p_{s,j}, p_{i,j}>$, as introduced above; $sl_j$ is a one-hot vector representing the corresponding human label; $\hat{p}_j=<\hat{p}_{e,j}, \hat{p}_{s,j}, \hat{p}_{i,j}>$ is the probability prediction of the current model; $\alpha$ is a hyper-parameter that balances the contributions of manual hard labels and self-distilled soft labels. We set $\alpha=0.5$ in all experiments for simplicity. $CE_{Loss}$ denotes the cross-entropy loss, which is standard for the hard label. For the soft label, the cross-entropy loss can be written as:

\begin{equation}
    CE_{Loss}(sl_j, \hat{p}_j) = - p_e * \log(\hat{p}_{e,j}) - p_s * \log(\hat{p}_{s,j}) - p_i * log(\hat{p}_{i,j})
\end{equation}

\subsection{Online Relevance Model}
\label{sec:online}

Based on the state-of-the-art CSRM-LLM (Stage-2), we distill a twin-tower model for online serving. The model architecture is similar to Que2Search~\cite{liu2021que2search} with XLM-Roberta-based tokenizer/encoder. The embedding layers are between the query and item towers, while other parameters are tuned separately. In the inference phase, the cosine similarity $c_{q, i}$ of the query and item towers' outputs will be used as the final prediction score. Note that $c_{q, i}$ ranges from $-1$ to $1$, which is not a valid probability applicable in computing KD loss. Following the strategy in ~\cite{liu2021que2search}, we set $\hat{p}_{q,i} =\sigma (s\cdot c_{q, i})$, where $\sigma$ denotes the Sigmoid function, and $s$ is a hyper-parameter set as $8$ empirically. We choose the twin-towered structure as it is more friendly for CPU serving. All product embeddings are pre-calculated and refreshed regularly. When receiving an online request, the query embedding is either generated in real time or directly fetched from online cache. 
For each query, we sample $M+N$ products to construct the query-product pairs for knowledge distillation, including: (1) Top $M$ impressed products from historical logs in recent one month. This data will guide the model to learn the relevance matching of popular query-product pairs. (2) Top $N$ products retrieved by the EBR model trained by the sparse click and conversion data collected in the new market. EBR candidates are included because the top impressed products are mostly relevant products, while we need more substitute and irrelevant items to balance the training data distribution.

Although the LLM is trained to predict three-class probabilities for each query-product pair, we distill two binary classifiers separately for online deployment. The exact match (EM) model identifies whether a product exactly matches the query, while the defect model detects irrelevant products. In the online pipeline, the defect model filters out irrelevant candidates, and the EM model provides a promotion score in the ranking stage. During knowledge distillation, we use cross-entropy loss between the model predictions and the soft labels generated by the teacher LLM. The loss function for the EM model is defined as follows:

\begin{equation}
    KD_{Loss} = \sum_{j \in \mathbf{D}} - (p_{e,j} * \log(\hat{p}_{e,j}) + (1 - p_{e,j}) * \log(1 - \hat{p}_{e,j}))
\end{equation}
where $p_{e,j}$ and $\hat{p}_{e,j}$ denote the predicted probabilities of exact match by the online model and teacher model, respectively, for the $j$-th sample in the training set $D$. We do not leverage hard labels here as they are very sparse compared to the soft labels. The knowledge distillation loss for the defect model can be formulated similarly. 

\section{Experiments}

\subsection{Dataset}
\subsubsection{Relevance Task}
Our dataset is collected exclusively from a production environment of a cross-market e-commerce company. We use South Korea as the source market, collecting 5 million relevance-labeled samples in Korean (KR) through human annotation. Chinese Taiwan serves as the target market, where we gather 500K samples, primarily in traditional Chinese (TC) with a small portion in English. In stage one, only KR data is used for training, while stage two incorporates both KR and TC data. We randomly select 20k samples from KR and TC respectively as the development set, while the rest ones are leveraged for training. For rigorous evaluation of the model's real-world effectiveness, we construct a high-quality test set of 20K samples across diverse product categories. Each sample is annotated by eight raters, with final labels retained only when at least seven raters agree.

\subsubsection{Translation Task}
We incorporate machine translation as auxiliary tasks, with training data categorized into four types: (1) Category mapping — 50K Korean (KR) category names manually mapped to traditional Chinese (TC); (2) Brand mapping — 50K KR brand names mapped to TC; (3) Title translation — 700K KR product titles translated into TC; and (4) Query translation — 30K selected KR queries translated into TC. All translations are human-curated to ensure quality and domain relevance.

\begin{table}[t]
\caption{Result Comparison}
\centering
\begin{tabular}{lccc}
\hline
\textbf{Model}   & \textbf{Setting} & \textbf{Training data} & \textbf{F1 score}  \\ \hline
GPT-4o   & Zero-shot & - & 0.850 \\ \hline
\multicolumn{4}{c}{\textbf{Stage-1: Cold-start with zero data}} \\ \hline
Roberta-base & SFT & KR & 0.764   \\
Roberta-large & SFT & KR & 0.780  \\ \hline
CSRM-LLM & SFT & KR & 0.908   \\
CSRM-LLM & SFT+MT & KR & 0.916   \\
CSRM-LLM & SFT+MT+RQA & KR & \textbf{0.937}   \\ \hline 
\multicolumn{4}{c}{\textbf{Stage-2: Localization by sparse data}} \\ \hline
CSRM-LLM & SFT+MT+RQA & KR+TC & 0.951 \\
CSRM-LLM & SFT+MT+RQA$^*$ & KR+TC & 0.956 \\
CSRM-LLM & SFT+MT+RQA$^*$+SD-1 & KR+TC & 0.962 \\
CSRM-LLM & SFT+MT+RQA$^*$+SD-2 & KR+TC & \textbf{0.965} \\ 
CSRM-LLM & SFT+MT+RQA$^*$+SD-3 & KR+TC & 0.964 \\ \hline 
\end{tabular}
\label{result_comp}
\end{table}

\subsection{Experimental Settings}
We conduct all experiments based on Orion-14B~\cite{chen2024orion}. At the beginning of this work, we compared different open-source multilingual large language models with similar sizes (including Llama3~\cite{touvron2023llama}, Mistral~\cite{jiang2023mistral}, Qwen2~\cite{yang2024qwen2} and Gemma2~\cite{lieberum2024gemma} etc.) and found that Orion performs better in both Korean and traditional Chinese for our scenario. 
To enable multi-task learning within a unified framework, we formulate relevance matching as a generation task. Given a query-product pair, the model generates one of three labels: 0 (irrelevant), 1 (substitute), or 2 (exact match). During training, only the conditional probabilities of these three tokens are normalized and included in the loss computation. We adopt the micro-averaged \textbf{F1 score} across the three classes as the evaluation metric.
\begin{equation}
    F1\_score = \frac{\sum_{i=0}^{2}TP_i}{\sum_{i=0}^{2}TP_i+\frac{1}{2}(\sum_{i=0}^{2}FP_i+\sum_{i=0}^{2}FN_i)}
\end{equation}
where $TP_i$, $FP_i$ and $FN_i$ are number of true positive, false positive and false negative samples for class $i$ ($i=0, 1, 2$), respectively. 

We use LoRA~\cite{hu2021lora} to speed up the fine-tuning procedures while preventing over-fitting. To be specific, we fine-tune only $q_{proj}$, $k_{proj}$, and $v_{proj}$ in each attention layer. The LoRA parameters are set as $r=128$, $\alpha=256$ and $drop\_out = 0.01$. The percentage of trainable parameters is 1.07\%. We use AdamW optimizer with initial learning rate $2e^{-5}$. We set the batch size as 64, maximum sequence length as 2048, and $L_2$ regularization as 0.01. The models are fine-tuned by a maximum of 500K steps, and the best checkpoint on the development set is selected as the final model.   
Detailed configurations of baselines and CSRM-LLMs are listed below.
\begin{itemize}
    \item \textbf{GPT-4o}~\cite{achiam2023gpt} is the best commercial LLM released by OpenAI. Due to its restriction on supervised fine-tuning, we adopt a zero-shot setting in our experiments.
    \item \textbf{RoBERTa-base/large}~\cite{DBLP:conf/acl/ConneauKGCWGGOZ20} is the multilingual version of the open-sourced XLM-RoBERTa-base/large model with supervised finetuning (SFT) on the relevance tasks. 
    \item \textbf{CSRM-LLM SFT} is the Orion-14b model with supervised fine-tuning on the relevance task.
    \item \textbf{CSRM-LLM SFT+MT} is the Orion-14b model, fine-tuned by the relevance task and machine translation (MT) task simultaneously. The relative weight of the auxiliary translation task is set as 0.1. 
    \item \textbf{CSRM-LLM SFT+MT+RQA} is the Orion-14b model jointly fine-tuned by the relevance and machine translation tasks, while Retrieval-based Query Augmentation (RQA) is incorporated into the prompt inputs.
    \item \textbf{CSRM-LLM SFT+MT+RQA+SD-$k$} is fine-tuned via $k$ rounds self-distillation (SD). Other settings remain the same as above. 
\end{itemize}

\subsection{Results}
The evaluation results of baseline models and different CSRM-LLM settings are summarized in Table \ref{result_comp}. As a strong baseline, GPT-4o obtains 85\% accuracy without additional fine-tuning, demonstrating its outstanding zero-shot text understanding capability. However, in the e-commerce scenario, there exist large amounts of domain knowledge and business-specific requirements, making it difficult for a zero-shot model to achieve the optimal performance. 

As baselines for stage one, the multi-lingual RoBERTa-base model fine-tuned by the Korean (KR) relevance data achieves only 0.764 F1 score on trainditional Chinese (TC) test set, while RoBERTa-large performs slightly better. After changing to the Orion backbone model with 14-billion parameters, the CSRM-LLM shows 12.8\% absolute lift compared to RoBERTa-large. This indicates that smaller language models have weaker cross-lingual generalization capability and a large model size is necessary in our scenario. The integration of translation tasks further boosts the F1 score from 0.908 to 0.916, which is statistically significant on the test set. Moreover, retrieval-based query augmentation improves the F1-score significantly to 0.937, demonstrating the necessity of injecting e-commerce knowledge on the query side. 

In stage two, about 500K relevance labels have been collected in traditional Chinese (TC), counting for 10\% of those in Korean (KR). After jointly training by KR and TC relevance labels, the F1 score is lifted by 1.4\% absolutely. Furthermore, by adding the sparse user conversion data in TC, we train a new EBR model for retrieval-based query augmentation (denoted as RQA$^*$). Replacing RQA with RQA$^*$ shows 0.5\% absolute enhancement, indicating the usefulness of local user behaviors.
Finally, we employ self-distillation training to mitigate the impact of labeling errors. Mechanistically, self-distillation acts as a form of label voting within the training set. We observe that the most significant improvement occurs after two rounds of self-distillation, achieving an F1 score of 0.965. Beyond this point, the performance stabilizes. Therefore, considering both cost and performance, a two-round self-distillation is optimal in our scenario. 
Detailed analyses of the proposed techniques will be provided in the later section.

\section{Analyses}

\subsection{How Machine Translation Tasks Help}
To prove that auxiliary translation tasks indeed activates the cross-lingual transfer ability of LLMs, we investigate the representation of data samples from the last hidden layer.
Intuitively, if a model has better cross-lingual capability, the representation of a query-product pair in different languages should be similar. Therefore, we measure \textit{cross-lingual transfer gap} based on a set of query-product pairs in both KR and TC languages:
\begin{equation}
    \frac{1}{m}\sum_{i=1}^{m}\frac{||e_i^{(2)} - e_i^{(1)}||}{||e_i^{(2)} - \overline{e}||}
\end{equation}
For the $i$-th query-product pair, the hidden representation in KR and TC are denoted by $e_i^{(1)}$ and $e_i^{(2)}$, respectively. $||e_i^{(2)} - e_i^{(1)}||$ is the $L_2$ distance between them. On top of that, we further normalize it by the distance between $e_i^{(2)}$ and $\overline{e}$, which is 
an average representation of all data samples in TC. This concludes the definition of the cross-lingual transfer gap on one single sample, and the average of all $m$ samples is taken as the overall metric.

We compute the \textit{cross-lingual transfer gap} for CSRM-LLM with and without auxiliary translation tasks. It turns out that the gap of LLM with the translation tasks (0.676) is smaller than that without translation tasks (0.689), which is consistent with our conjecture. We also examine the samples which turns to be correct after adding the auxiliary translation task in training. A concrete example is query = ``韓國菊花茶" (``Korean chrysanthemum tea") and product title = ``epanie 涼茶花茶" (``epanie herbal tea") in traditional Chinese. The correct label is substitute while it is labeled as irrelevant without trained by translation tasks. We found there is a similar sample in the Korean training dataset: query = ``\begin{CJK*}{UTF8}{mj}에케네시아차\end{CJK*}" (``Echinacea tea"), product title = ``\begin{CJK*}{UTF8}{mj}에빠니 허브뜨레 꽃차\end{CJK*}" (epanie herbal tea), and label = ``substitute". The normalized distance of these two pairs is shortened from 0.735 to 0.495 after adding translation tasks. This shows that the original model did not transfer cross-lingual knowledge well, but after the translation tasks are added, the LLM effectively captures cross-lingual semantics and generalizes the labeling guideline to similar samples. 

\subsection{Retrieval-based Query Augmentation}
First, we show a motivating example for retrieval-based query augmentation (RQA). The query-product pair is labeled as substitute: query = ``安城" (Anseong, name of the city famous for producing traditional noodle soup) and product title = ``OTTOGI 不倒翁 辣牛肉湯麵 增量版, 104g, 6入" (OTTOGI Tumbler Spicy Beef Noodle Soup, Increasing Version, 104g, 6 pieces). Without RQA, the LLM prediction is irrelevant since it cannot understand the intent of the query. With RQA, the retrieved text is ``Nongshim 農心 安城湯麵 125g" (Nongshim Anseong Noodle Soup) which indicates that the query intent is to find Anseong noodle soups. Therefore, the RQA-enhanced LLM can correctly predict the sample as substitute.

To compare the quality of different versions of RQA features, we ask human raters to examine the accuracy of retrieved texts on a random subset. Each retrieved text is labeled as correct if it captures the semantics of the query and does not introduce additional noise. The RQA features directly generated by the BM25 algorithm hold only 65\% accuracy, while the EBR model distilled by LLM rankers improves the accuracy to 82\%. After adding local behavior data to train the EBR model, the RQA features for stage two (denoted by RQA$^*$) achieve an accuracy of 88\%.

\subsection{Effectiveness of Self-Distillation}
\begin{figure}[h]
  \centering
  \includegraphics[width=1.0\linewidth]{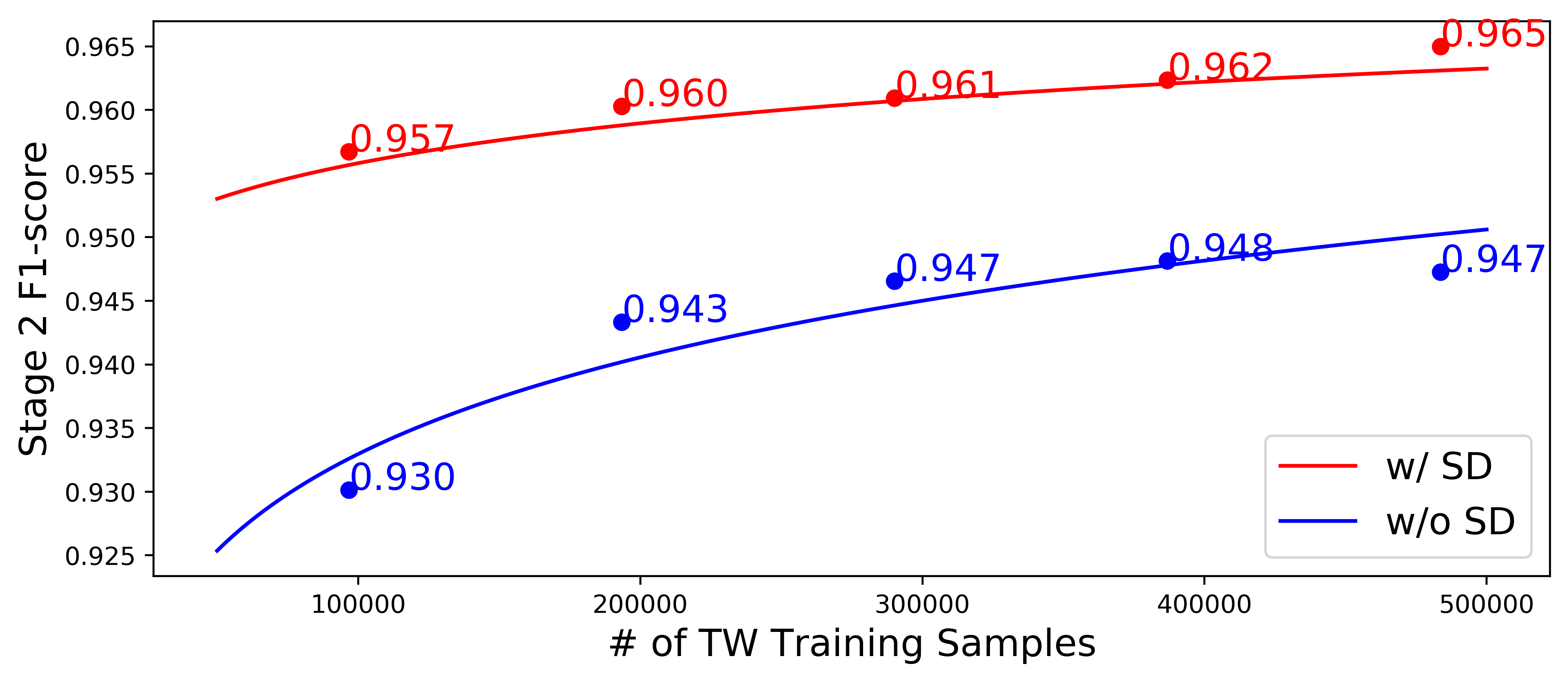}
  \caption{F1 score vs. number of relevance training samples in local market, w/ and w/o self-distillation. 
  }
  \label{fig:dfc_acc_vs_tr_smpl}
\end{figure}

Figure \ref{fig:dfc_acc_vs_tr_smpl} illustrates the impact of self-distillation while varying the local training dataset size. As shown in the figure, the CSRM-LLM with self-distillation training (indicated by the red line) consistently outperforms the counterpart without self-distillation (the blue line). The impact of labeling noise is large when the local training dataset is small, which may be caused by severe over-fitting. 
This demonstrates that the self-distillation technique indeed assists LLMs to minimize the impact of labeling errors, which are inevitably introduced by human annotators.

To show the effectiveness of self-distillation with a concrete example, here is a sample where query = ``孕婦" (pregnant woman) and  product title = ``達摩本草 孕哺媽咪卵磷脂粉包 30包, 105g, 1盒" （Damo Herbal Lecithin Powder Packets for Pregnant and Breastfeeding Moms 30 packs, 105g, 1 box). Clearly it should be relevant, but was rated as irrelevant by mistake. During self-distillation, LLM rates this sample as 77.77\% relevant, 17.90\% substitute and 4.32\% irrelevant. Consequentially, self-distillation alleviates the label noise in the next round of training.

\section{Production Impacts}
Based on CSRM-LLM’s relevance predictions, we distill an online model for deployment, as described in Section~\ref{sec:online}. 
As both offline evaluation and online evaluation show promising results, we have launched the model to full traffic in the production. This launch is crucial for us to build customers' trust on our e-commerce platform by displaying more relevant items to users' queries at top positions.

\subsection{Offline Evaluation}
As introduced in Section \ref{sec:online}, the online model has a twin-tower architecture, taking query and product titles as inputs to each tower separately. Each tower is a pretrained XLM-Roberta-Base model \cite{DBLP:conf/acl/ConneauKGCWGGOZ20}, which shares the embedding layer and maintains separate parameters for the rest layers. In each tower, the last hidden layer of the encoder's output is average pooled and  projected to $128$ dimension by a linear layer to form the final output embedding. 
Following the logic described in Section \ref{sec:online}, we collect training samples for knowledge distillation with impressed products ($M=20$) and EBR products ($N=100$ ) for each query. 
The batch size is set as $1024$. We adopt a linear decay learning rate scheduler which starts from $8e^{-5}$ and ends at $1e^{-6}$. The online baseline models also adopt the same twin-tower architecture, but are trained directly by the hard relevance labels without knowledge distillation.

Table \ref{distill_offline} shows the offline evaluation metrics, where we report the binary F1 and accuracy metrics for the exact match model and defect model separately. It turns out that the defect model's performance is very close to the teacher LLM, which significantly outperforms the original online baseline. The exact match model has a slightly larger gap. We can further close the gap by replacing the twin-tower architecture by a single-tower architecture that takes the concatenation of queries and product titles as inputs. We leave this as future work as it requires modifications to the current online stacks for GPU serving.

\begin{table}[t]
\caption{Knowledge Distillation Offline Metrics}
\centering
\begin{tabular}{lccc}
\hline
Task   & Model & Accuracy & F1  \\ \hline
{} & Online baseline & 0.7420 & 0.7050 \\
\cline{2-4}
Defect  & Offline CSRM-LLM & 0.9680 & 0.9518 \\ 
\cline{2-4}
{} & Distilled online model & 0.9675 & 0.9509 \\ \hline
{} & Online baseline & 0.8290 & 0.8150 \\
\cline{2-4}
Exact Match & Offline CSRM-LLM & 0.9673 & 0.9553 \\ 
\cline{2-4}
{} & Distilled online model & 0.9504 & 0.9317 \\ \hline
\end{tabular}
\label{distill_offline}
\end{table}

\subsection{Online Evaluation}

During online evaluation, half users are served by the original online baseline (Treatment A), and others are served by the new model distilled from the CSRM-LLM (Treatment B). First, we ask human raters to judge the search results on a random subset of the user queries. We leverage EM@1 and Defect@5 to measure the relevance performance. The EM@1 metric reveals the fraction of queries that can display exact match items at the topmost position, whereas the Defect@5 metric measures the proportion of queries that do not exhibit defective items within the top five positions. The result suggests a relative 3.5\% EM@1 improvement and $45.8\%$ Defect@5 reduction by comparing treatment A and B.

We observed several successful cases in our online experiments. For example, when a user searched for ``lake salt", the baseline model ranked a potato chip with lake salt flavor among the top results, an irrelevant item given the user's intent. In contrast, the distilled model correctly prioritized products such as natural lake salt for cooking, which better align with the query intent. Similarly, for the query ``paint", the baseline surfaced painting tools, whereas the distilled model ranked actual paint products such as wall paint, providing more accurate and useful results.

Next, we start a live A/B test to compare the commercial metrics of these two treatments. The online results suggest a statistically significant $+0.866\%$ increase on session conversion rate\footnote{session conversion rate is defined as the portion of search sessions with customer adding-to-cart or direct purchase.}. A session generally describes one customer stay, typically lasting as long as $30$ minutes and possibly consists of multiple queries. A typical customer's experience is: after she/he issues a query, the returned search results match the query's intent but the customer wants to do more exploration via other follow-up queries. Better relevance results will boost the customers' trust to the platform and promote user engagement. Finally, the customer finds desired products and makes conversions within the same session. Based on the positive metrics for both relevance score and customer feedback, we launched the distilled model to the production mainstream, while the hold-out traffic further demonstrated a consistent enhancement.

\section{Related Work}
\textbf{Relevance matching models} have been widely studied in multiple areas such as web search and online e-commerce (\cite{DBLP:journals/ftir/XuHL20}), since the beginning of the deep learning era. Early work primarily explored the importance of different types of information and signals for relevance-matching tasks, such as the textual matching between queries and documents and user click behavior. DSSM \cite{DBLP:conf/cikm/HuangHGDAH13} uses DNNs to rank documents by calculating the similarity between queries and documents in vector space. DRMM \cite{DBLP:conf/ccir/YangLGFZLWC18} focused more on interaction information to consider the factors of exact matching signals, query term importance, and diverse matching requirements. Subsequent research has delved into the impact of various network architectures on this issue, including the use of CNNs \cite{DBLP:conf/cikm/PangLGXXC17,DBLP:conf/wsdm/DaiXC018,DBLP:conf/kdd/ChenHNLLWX18}, GRUs \cite{DBLP:conf/cikm/PangLGXXC17}, Attention \cite{DBLP:conf/cikm/ChenCXLZ23}, and their variants \cite{DBLP:conf/sigir/XiongDCLP17,DBLP:conf/kdd/ChenHNLLWX18}.

\textbf{Large language models (LLMs) for relevance matching} have become a new research focus with the advancement of LLMs.  The most straightforward way to utilize LLMs to facilitate relevance matching is data/label augmentation. Thomas et al.~\cite{thomas2024large} proposes a method to use LLMs to label whether a document is useful for a given search. Liu et al. \cite{DBLP:journals/corr/abs-2404-02616} use LLMs to augment search relevance data, by letting LLMs to rewrite a small but high-quality batch of query document pairs. Another line of works utilize LLM directly as the relevance marker. Liu et al. \cite{DBLP:journals/corr/abs-2304-10149} conduct a preliminary study on whether LLMs (ChatGPT in this case) are good relevance matching models. According to the experiment results, we can learn that LLMs perform well, especially with few-shot prompting. He et al. \cite{DBLP:conf/cikm/HeXJSLFMKM23} conducted a similar study to test whether LLMs can serve as conversational recommenders, also through standard prompt engineering methods. The results also support that LLMs can perform well in relevance matching tasks.
\section{Conclusion}
In this paper, we introduce the methodology of Coupang to roll out relevance models to emerging e-commerce markets. A multi-task multi-lingual LLM-based relevance matching framework with self-distillation and retrieval-based query augmentation is proposed for tackling the cold-start challenges. Our experiments demonstrate that the proposed framework significantly advances the SOTA, improving performance by over 10\% compared to directly using GPT-4o. After applying knowledge distillation, the online model meets the deployment requirements with nearly no loss in effectiveness, achieving 45.8\% reduction in defect ratio and 0.866\% enhancement in session purchase rate. 
\section{GenAI Usage Disclosure}
Generative AI tools were used solely for minor language editing, such as grammar and clarity improvements, comparable to standard writing assistants like Grammarly. No content was generated or substantially rewritten using generative AI software.


\bibliographystyle{ACM-Reference-Format}
\balance
\bibliography{reference_final}

\end{CJK*}
\end{document}